\documentclass[twocolumn]{aastex631}

\usepackage{xcolor}
\usepackage{graphicx}
\usepackage{amsmath}
\usepackage{natbib}

\def\isothermal{
\begin{figure*}[ht!]
\hspace*{-2cm}\includegraphics[width=1.23\textwidth]{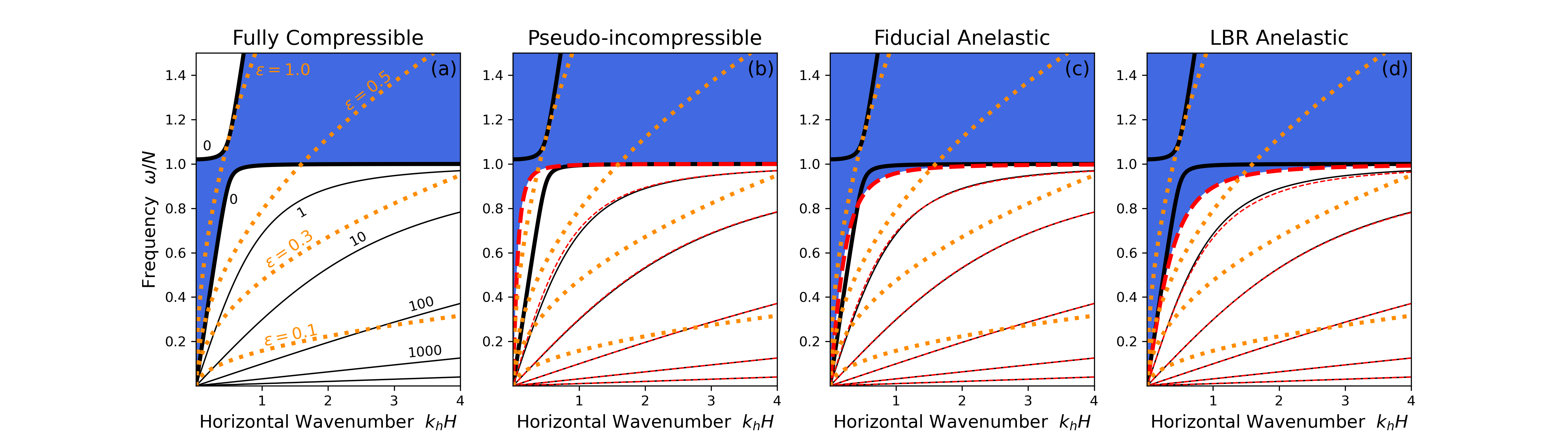}
\caption{\footnotesize  Propagation diagrams for an isothermal atmosphere for four treatments of the fluid equations: (a) a fully compressible fluid---i.e., no approximation, (b) the pseudo-incompressible condition, (c) the fiducial anelastic approximation, and (d) the LBR formulation of the anelastic approximation (see Table~\ref{tab:table1} for a summary). In each panel, the solid black curves correspond to the isocontours of the square of the dimensionless vertical wavenumber $(k_z H)^2$ for a fully compressible atmosphere (where the density scale height $H$ is a constant function of height for an isothermal atmosphere). The value of each contour is indicated by a black label in panel a. Further, the thick black contour corresponds to the zero contour that separates domains of vertical wave propagation ($k_z^2 > 0$) and evanescence ($k_z^2 < 0$). In panels $b$--$d$, the dashed red curves indicate the same contours but for the approximation indicated at the top of the panel. In each panel, the domain of evanescent waves is indicated by the blue shading, while the region of vertical propagation is unshaded. The dotted curves in each panel are isocontours of the dimensionless frequency.  Since the dimensionless frequency is a function of wavenumber, $\eps = \omega/\sqrt{gk_h}$, isocontours are curved lines with low values in the lower-right portion of the diagram and high values in the upper left. All approximations reproduce the correct vertical wavenumber when the dimensionless frequency $\eps$ is small. Differences between the approximations begin to appear for moderate to large values of the dimensionless frequency $\eps > 0.3$.} 
\label{fig:isothermal}
\end{figure*}
}

\def\polytrope{
\begin{figure*}[ht!]
\includegraphics[width=1.0\textwidth]{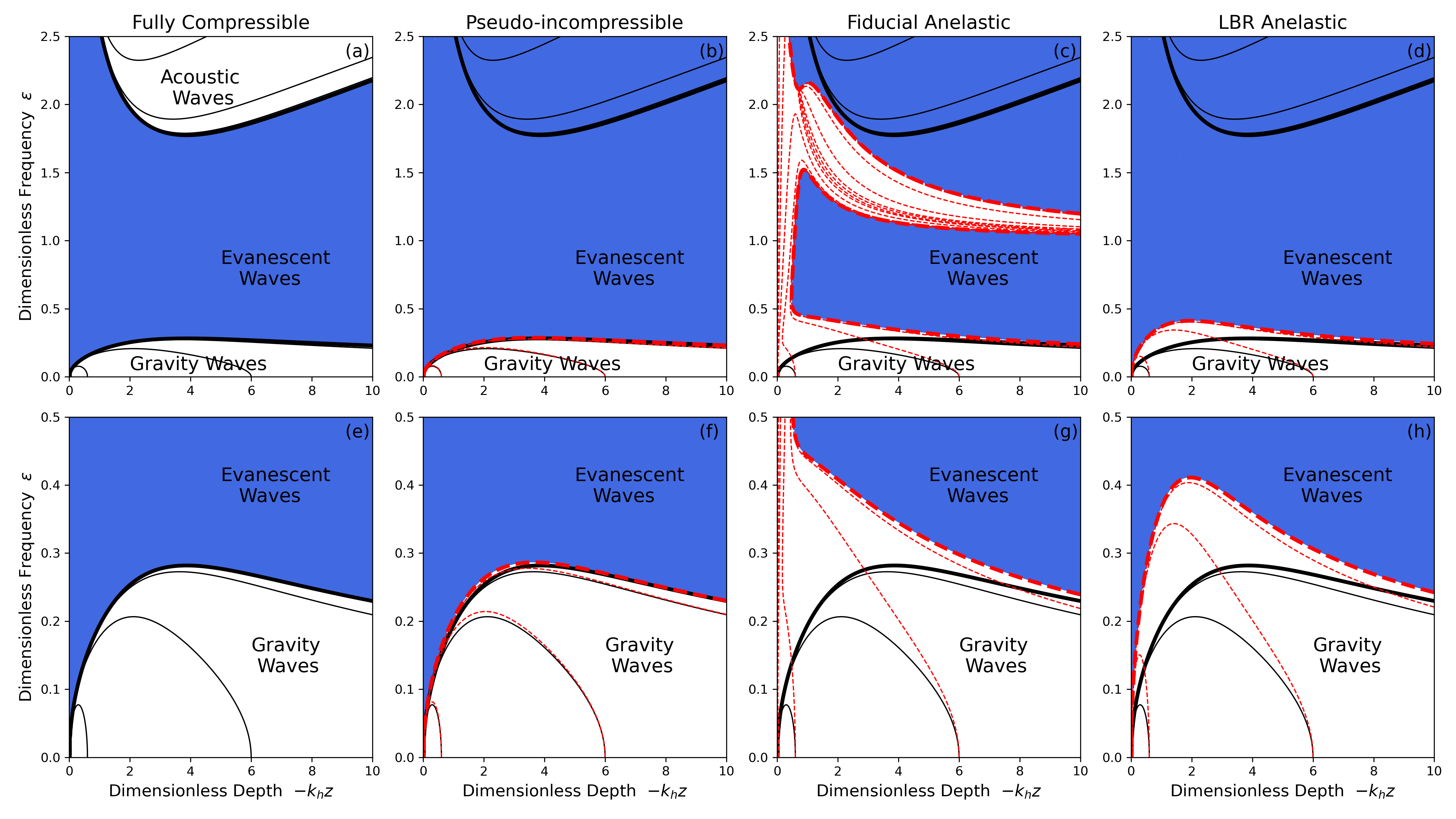}
\caption{\footnotesize  Propagation diagrams for a polytropic atmosphere under different approximations to the fluid equations. In each panel, the solid black curves correspond to the isocontours of the square of the dimensionless vertical wavenumber $(k_z /k_h)^2$ for a fully compressible atmosphere.  These contours are plotted versus a non-dimensional depth, $-k_h z$, and the dimensionless frequency, $\eps = \omega/\sqrt{gk_h}$. The thick black contour corresponds to the zero contour that separates domains of vertical propagation ($k_z^2 > 0$) and evanescence ($k_z^2 < 0$). The dashed red curves indicate the same contours but for the approximation indicated at the top of the column. The background colors have the same meaning as in Figure~\ref{fig:isothermal}. The upper panels illustrate a larger range of frequency and capture the high-frequency acoustic branch. The pseudo-incompressible and LBR anelastic approximations eliminate all such acoustic waves. The fiducial anelastic approximation leaves a highly distorted residual domain of propagating acoustic waves. In general, all three approximations do well in reproducing the correct vertical wavenumber when the dimensionless frequency is small $\eps \lesssim 0.1$.  However, the pseudo-anelastic approximation has the least distortion to the spatial extent of the wave cavity even for frequencies as large as $\eps\approx 0.3$.}
\label{fig:polytrope}
\end{figure*}
}

\def\tableone{
\begin{table*}[h!]
  \begin{center}
    \textbf{Table~\ref{tab:table1}}\\
    {Comparison of Sound-Proofing Techniques}\\
    \begin{tabular}{|l|c|l|c|}
      \hline
       &  &  & \textbf{Envelope} \\
      \textbf{Equation Set} & $\alpha(z)$ & \textbf{Square of the Vertical Wavenumber} $k_z^2$ & \textbf{Scale}  $\Lambda$ \\
      \hline\hline
      Fully-Compressible & $g^2k_h^2 - \omega^4$ &  $\D k_h^2 \left({\frac{N^2}{\omega^2}}-1\right)-\frac{\omega_c^2}{c^2} + {\rm O}(\eps^2)$ &  $H$ \\
      \hline
      Pseudo-incompressible &  $\D g^2k_h^2-\omega^4 +\omega^2 \frac{g}{H_*}$ &  $\D k_h^2 \left({\frac{N^2}{\omega^2}}-1\right) -\frac{\omega_c^2}{c^2} - \frac{N^2}{c^2}+ {\rm O}(\eps^2)$ &  $H + {\rm O}(\eps^2)$ \\
      \hline
      Fiducial Anelastic &  $\D g^2k_h^2-\omega^4 +\omega^2 \frac{g}{H}$ & $\D k_h^2 \left({\frac{N^2}{\omega^2}}-1\right) -\frac{\omega_c^2}{c^2} + \frac{N^2}{4g} \frac{H+H_*}{HH_*} + \frac{1}{2g}\frac{dN^2}{dz} + {\rm O}(\eps^2)$ & $H_* + {\rm O}(\eps^2)$ \\
      \hline
      LBR Anelastic & $\D g^2k_h^2-\omega^4 +\omega^2 \left(N^2 + \frac{g}{H}\right)$ & $\D k_h^2 \left({\frac{N^2}{\omega^2}}-1\right) -\frac{\omega_c^2}{c^2} - \frac{1}{g} \frac{dN^2}{dz} + {\rm O}(\eps^2)$ & $H + {\rm O}(\eps^2)$ \\
      \hline
    \end{tabular}
  \end{center}
  \caption{Wave properties achieved under various sound-proofing approximations as indicated in the first column. The second column indicates the function $\alpha(z)$. The third and fourth columns provide the square of the local vertical wavenumber $k_z^2$, and the scale length $\Lambda$ of the amplitude envelope for internal gravity waves in the low-frequency limit. The wave frequency and horizontal wavenumber are indicated by $\omega$ and $k_h$, respectively. The atmosphere is characterized by the vertical profiles of the sound speed $c$, the density scale height $H$, the scale height for an adiabatic stratification (i.e., the scale height for the potential density) $H_* = c^2/g$, the buoyancy frequency $N$, and the acoustic cutoff frequency $\omega_c$. For the vertical wavenumber and envelope scale, all terms with a magnitude ${\rm O}(\eps^2)$ or smaller have been neglected. Since, the leading-order terms in the vertical wavenumber are ${\rm O}(\eps^{-2})$, the neglected terms are small by a factor of $\eps^4$, i.e., they are fourth order.    \label{tab:table1}}
\end{table*}
}

\def\tabletwo{
\begin{table*}[h!]
  \begin{center}
    \textbf{Table~\ref{tab:table2}}\\
    {Fractional Errors Introduced by Sound-Proofing Techniques}\\
    \begin{tabular}{|l|c|c|c|}
      \hline
        & \textbf{Errors in the} &  \textbf{Errors in the} & \textbf{Errors in the} \\
      \textbf{Equation Set} & \textbf{Vertical Wavenumber} $k_z$ & \textbf{Envelope Scale}  $\Lambda$ & \textbf{Wave Functions} $\delta P$, $u$\\
      \hline\hline
      Pseudo-incompressible & ${\rm O}(\eps^2)$ & ${\rm O}(\eps^2)$ & ${\rm O}(\eps^2)$, ${\rm O}(\eps^2)$ \\
      \hline
      Fiducial Anelastic & ${\rm O}(\eps^2)$ & ${\rm O}(1)$ &  ${\rm O}(\eps)$,  ${\rm O}(\eps)$ \\
      \hline
      LBR Anelastic & ${\rm O}(\eps^2)$ & ${\rm O}(\eps^2)$ & ${\rm O}(\eps^2)$, ${\rm O}(\eps)$ \\
      \hline
    \end{tabular}
  \end{center}
  \caption{Magnitude of the fractional errors that are introduced in internal gravity waves by three different sound-proofing techniques. Each column lists the size of the error divided by the leading order behavior for the wave property indicated at the top of the column. The size of each error is presented in terms of the dimensionless frequency $\eps = \omega/\sqrt{gk_h}$. The pseudo-incompressible approximation evinces the smallest errors, all appearing at second order. Both of the anelastic approximations have errors that appear at first order or larger.\label{tab:table2}}
\end{table*}
}

\newcommand{\bvec}[1]{{\mbox{{\boldmath$#1$}}}}	
\newcommand{\unitv}[1]{\bvec{\hat{#1}}}			
\newcommand{\grad}{\bvec{\nabla}}			    

\newcommand{\dP}{\delta P}			
\newcommand{\eps}{\varepsilon}

\newcommand{\D}{\displaystyle}
\newcommand{\eqnref}[1]{(\ref{#1})}

\begin{document}

\title{Low-Frequency Internal Gravity Waves are Pseudo-incompressible}

\author{Bradley W. Hindman}
\affil{JILA, University of Colorado, Boulder, CO~80309-0440, USA}
\affil{Department of Applied Mathematics, University of Colorado, Boulder, CO~80309-0526, USA}

\author{Keith Julien}
\affil{Department of Applied Mathematics, University of Colorado, Boulder, CO~80309-0526, USA}


\begin{abstract}

Starting from the fully compressible fluid equations in a plane-parallel atmosphere, we demonstrate that linear internal gravity waves are naturally pseudo-incompressible in the limit that the wave frequency $\omega$ is much less than that of surface gravity waves, i.e., $\omega \ll \sqrt{g k_h}$ where $g$ is the gravitational acceleration and $k_h$ is the horizontal wavenumber. We accomplish this by performing a formal expansion of the wave functions and the local dispersion relation in terms of a dimensionless frequency $\eps = \omega / \sqrt{gk_h}$. Further, we show that in this same low-frequency limit, several forms of the anelastic approximation, including the Lantz-Braginsky-Roberts (LBR) formulation, poorly reproduce the correct behavior of internal gravity waves. The pseudo-incompressible approximation is achieved by assuming that Eulerian fluctuations of the pressure are small in the continuity equation. Whereas, in the anelastic approximation Eulerian density fluctuations are ignored. In an adiabatic stratification, such as occurs in a convection zone, the two approximations become identical.  But, in a stable stratification, the differences between the two approximations are stark and only the pseudo-incompressible approximation remains valid.

\end{abstract}



\section{Introduction}
\label{sec:Introduction}

Numerical simulations of convection in low-mass stars, the Earth's atmosphere, giant planets, and many other astrophysical objects all must face the tyranny of sound.  Generally, sound waves propagate quickly and have high frequencies; thus, the typical timescale associated with acoustics is far shorter than those arising from convection and large-scale circulations. In a numerical simulation, this short timescale ensures through the CFL condition that sound waves control the size of the timestep that can be taken while still maintaining numerical stability. The difference can be dramatic.  For example, at the base of the Sun's convection zone, the speed of sound is roughly 200 ${\rm km \, s}^{-1}$ while the convective flow speed is on the order of 20 ${\rm m\,s}^{-1}$ \citep[e.g.][]{Miesch:2012}. A numerical simulation that is forced to track sound waves for stability will need to take $10^4$ times as many time steps to evolve the solution for the same duration as a simulation that could ignore the acoustic wave field.  This inflation of the necessary computational work is particular onerous since the immense timescale difference between the deep convection and the sound waves indicates that the two phenomena are essentially decoupled.

A variety of methods have been proposed to mitigate this dilemma; almost all involve modifications to the fluid equations to either temper the impact of sound waves or to remove sound altogether. One way to reduce the influence of sound on the time step is to artificially lower the speed at which sound waves propagate \citep[e.g.,][]{Rempel:2005, Rempel:2006, Hotta:2012, Kapyla:2016, Ijima:2019}. Successful application of such Reduced Speed of Sound Techniques (RSST) requires that the sound speed be reduced sufficiently to make sound waves tractable, but to maintain enough celerity in the sound waves such that they do not interact strongly with the convective motions.

A more common solution is to surgically remove terms from the continuity equation such that sound waves are no longer a permissible solution to the fluid equations. These ``sound-proofed" equation sets typically apply to low-Mach number motions with small thermodynamic fluctuations about a hydrostatic background atmosphere. The most venerable of these techniques is the Boussinesq approximation, whereby the fluid is assumed to be incompressible with constant density. In the highly stratified atmospheres of stars and giant planets where the mass density can vary by orders of magnitude, treatments that can account for the stratification are necessary. In these stratified systems, the fundamental presumption is that for sedate motions a displaced parcel of fluid quickly equilibrates thermodynamically with its new surroundings. In astrophysics the most common of these extensions to the Boussinesq framework is the anelastic approximation \citep[e.g.,][]{Batchelor:1953, Ogura:1962, Gough:1969, Gilman:1981,bannon1996anelastic}, which removes all density fluctuations that appear in the continuity equation. A similar technique called the pseudo-incompressible approximation is a bit subtler, removing only the influence of Eulerian pressure fluctuations from the continuity equation \citep[e.g.,][]{Durran:1989, klein2009asymptotics, Vasil:2013}.

Such sound-proofing techniques have been used extensively in stellar and planetary convection simulations where the convecting layer spans many density scale heights. In regions of efficient convection, where the redistribution of heat and mass by the convective motions efficiently drives the atmosphere towards an adiabatic stratification, the most common forms of the anelastic and pseudo-incompressible equations are identical and either approximation works well.  However, in a stably stratified fluid, the two approximations differ to the extent that they may violate their underlying assumptions, leading to different dynamics. Specifically, \citet{klein2010regime}, \citet{Brown:2012} and \citet{Vasil:2013} have demonstrated that anelastic formulations do a disservice to internal gravity waves leading to a loss of energy conservation and to large errors in the wave frequencies.  Further, \citet{klein2010regime} and \citet{Vasil:2013} have demonstrated that although the pseudo-incompressible approximation does far better in preserving the properties of internal gravity waves, it too evinces discrepancies from the fully compressible wave forms.

Here, we demonstrate that internal gravity waves naturally approach the pseudo-incompressible limit as their frequency becomes very low. The discrepancies noted by \citet{klein2010regime} and \citet{Vasil:2013} arise only when the wave frequencies become large and the assumption of sedate motions in a state of pressure-equilibrium is lost. We accomplish this by deriving internal gravity waves in a plane-parallel atmosphere with a general stratification and subsequently performing a low-frequency expansion of the local dispersion relation and of the wave functions. We find that, to lowest-order in the frequency, internal gravity waves are incompressive.  To the next order in the frequency, they become pseudo-incompressible. All forms of the anelastic approximation fail to produce the correct behavior for both the dispersion relation and the wave functions.  

In the next section we formulate the anelastic and pseudo-incompressible approximations.  Section~\ref{sec:IGW} derives the governing equation for internal gravity waves in a general stratification for a fully compressible fluid. We explore the low-frequency limit of these waves in Section~\ref{sec:low-frequency_limit}, deriving the magnitude and ordering of terms in the continuity and momentum equations.  In Section~\ref{sec:sound-proofing} we rederive internal gravity waves using three different sound-proofed equation sets and discuss the integrity of each approximation.  Finally, in Section~\ref{sec:discussion} we summarize and discuss the implications of our results.

\section{Sound-Proofing Formulations}
\label{sec:Formulations}


\subsection{The Anelastic Approximation}
\label{subsec:intro_Anelastic}

The anelastic condition is a relatively simple replacement for the continuity equation that captures significant density variation in the mean properties of the fluid.  For instance, in a gravitationally stratified fluid with velocity, $\bvec{u}$, and time-averaged density that varies with height, $\rho_0(z)$, the continuity equation is replaced with

\begin{equation}
    \label{eqn:anelasticity}
    \grad\cdot\left(\rho_0 \bvec{u}\right) = 0 \; .
\end{equation}

\noindent This expression can be derived from the full continuity equation,

\begin{equation}
    \label{eqn:full_continuity}
    \frac{\partial \rho}{\partial t} + \grad \cdot \left(\rho \bvec{u}\right) = 0\; ,
\end{equation}

\noindent by making two assumptions that are often appropriate for flows of low Mach number: 1) the time derivative of the mass density $\rho$ is inconsequential and 2) the fractional fluctuations of the density around the background density are small, i.e., $\vert \rho_1/\rho_0 \vert \ll 1$ where $\rho = \rho_0 + \rho_1$. The popularity of the anelastic approximation arises from two important properties. When the continuity equation is replaced by the anelastic condition, Equation~\eqnref{eqn:anelasticity}, sound waves are removed as a permissible solution to the fluid equations and the mass flux $\rho_0 \bvec{u}$ can be written using stream functions.

\citet{Brown:2012} and \citet{Vasil:2013} both remarked that when the anelastic form of the continuity equation is employed, the fluid equations are no longer energy conserving without modifications to the momentum equation. To enforce conservation of energy, an otherwise unmotivated change to the buoyancy force is required. For an inviscid fluid, the vertical momentum equation can be written in the following form, 

\begin{equation}
    \label{eqn:anelastic_momentum}
    \rho_0 \frac{Dw}{Dt} = -\rho_0 \frac{d}{dz} \left(\frac{P_1}{\rho_0}\right) + \frac{g \rho_0}{c_p} s_1 + \frac{N^2}{g} P_1 \; ,  
\end{equation}

\noindent with the pressure $P$ and specific entropy density $s$ decomposed into a steady hydrostatic background and a fluctuation about that background, $P = P_0 + P_1$ and $s = s_0+s_1$. The vertical velocity is $w$, $c_p$ is the specific heat capacity at constant pressure, and $z$ is the height within the atmosphere with concomitant unit vector $\unitv{z}$ anti-aligned with gravity, $\bvec{g} = -g \unitv{z}$. Further, the quantity $N^2=g c_p^{-1} ds_0/dz$ is the square of the atmosphere's buoyancy or Brunt-V\"ais\"al\"a frequency. In Equation~\eqnref{eqn:anelastic_momentum}, we have ignored the density fluctuation in the inertial term on the left-hand side, subtracted the steady hydrostatic component from the force balance, and used the ideal gas law to rewrite the density fluctuation in terms of the pressure and entropy fluctuations. To ensure energy conservation, the term involving the buoyancy frequency must be discarded or be physically subdominant. In a convection zone, where efficient heat transport drives the atmosphere towards an adiabatic gradient with $N^2\approx 0$, this approximation is completely justified and has been coined the Lantz-Braginsky-Roberts (LBR) formulation of the anelastic approximation \citep{Lantz1992, BR1995}. Conversely, in a stably stratified region, the term is not small and cannot generally be self-consistently ignored.

We will examine two distinct formulations of the anelastic approximation. Both replace the continuity equation with the anelastic condition~\eqnref{eqn:anelasticity}.  One of these approximations---which we will dub the ``fiducial" anelastic approximation---will make no further assumptions, leaving the momentum equation unmodified. The other formulation will be the LBR anelastic approximation as discussed above, which ensures energy conservation by excising a specific term from the momentum equation.


\subsection{The Pseudo-Incompressible Approximation}
\label{subsec:intro_PI}

The pseudo-incompressible approximation as proposed by \citet{Durran:1989} modifies the continuity equation under the assumption that Eulerian fluctuations of the gas pressure can be ignored.  Following \citet{Durran:2008}, we start by defining the potential density $\rho_*$ for an ideal gas,

\begin{eqnarray}
    \label{eqn:def_potential_density}
    \rho_* \equiv \rho \, e^{s/c_p} \; .
\end{eqnarray}

\noindent If we take the convective derivative of the potential density and utilize the continuity equation~\eqnref{eqn:full_continuity} and the thermal energy equation,

\begin{equation}
    \label{eqn:full_energy}
    \rho T \frac{D s}{Dt} = Q \; ,
\end{equation}

\noindent we obtain a prognostic equation for the potential density

\begin{equation}
    \frac{1}{\rho_*} \left(\frac{\partial\rho_*}{\partial t} + \bvec{u}\cdot\grad\rho_*\right) = -\grad\cdot\bvec{u} + \frac{Q}{c_p\rho T} \; ,
\end{equation}

\noindent where $T$ is the temperature and $Q$ represents all irreversible thermodynamic processes, such as thermal diffusion, viscous heating, radiative transfer, etc. Finally, by invoking Equation~\eqnref{eqn:def_potential_density} and the equation of state for an ideal gas,

\begin{equation}
    \frac{1}{\rho_*}\frac{\partial \rho_*}{\partial t} = \frac{1}{\rho}\frac{\partial \rho}{\partial t} + \frac{1}{c_p} \frac{\partial s}{\partial t} = \frac{1}{\gamma P} \frac{\partial P}{\partial t} \; ,
\end{equation}

\noindent we replace the time derivative of the potential density with the time derivative of the gas pressure,

\begin{equation}
    \label{eqn:alt_continuity}
    \grad \cdot \left(\rho_* \bvec{u}\right) = \frac{\rho_*}{\rho} \left(\frac{Q}{c_p T} -\frac{1}{c^2}\frac{\partial P}{\partial t} \right) \; .
\end{equation}

\noindent In the preceding equations, $\gamma$ is the gas's adiabatic exponent and $c$ is the sound speed given by $c^2 = \gamma P/\rho$.

Equation~\eqnref{eqn:alt_continuity} is an exact form of the continuity equation for which no approximation has been made other than the gas being ideal. The pseudo-incompressible approximation is achieved by assuming that the term involving the time derivative of the gas pressure is negligible,

\begin{equation}
    \label{eqn:alt_continuity2}
    \grad \cdot \left(\rho_* \bvec{u}\right) = \frac{\rho_*}{\rho} \frac{Q}{c_p T} \; .
\end{equation}   

\noindent Such an approximation is valid in the limit of infinite sound speed and is consistent with slow motions of low Mach number for which a displaced parcel of fluid rapidly reaches pressure equilibration with its new surroundings. Most importantly, making this approximation removes sound waves from the fluid equations in the same way that anelasticity does. Durran's form of the pseudo-incompressible approximation \citep{Durran:2008} involves replacing the continuity equation by the preceding equation, but otherwise leaving the other fluid equations unmodified---specifically, the momentum equation remains the same. 

For isentropic motion, the pseudo-incompressible condition reduces to a form that is reminiscent of the anelastic relation

\begin{equation}
    \label{eqn:PI_continuity}
    \grad\cdot \left(\rho_* \bvec{u}\right)=0 \; ,
\end{equation}

\noindent with the mass density replaced by the potential density. However, for flows with low Mach number, thermodynamic fluctuations are small and we can safely linearize Equation~\eqnref{eqn:PI_continuity}, replacing the potential density by the potential density of the hydrostatic background atmosphere (denoted by `0' subscripts),

\begin{equation}
    \label{eqn:ref_potential_density}
    \rho_{*0} \approx \rho_0 e^{s_0/c_p} = \left(\frac{\hat{\rho}}{\hat{P}^{1/\gamma}}\right) \, P_0^{1/\gamma} \; .
\end{equation}

\noindent The last equivalency in Equation~\eqnref{eqn:ref_potential_density} arises by noting that the potential density is the density that a fluid parcel would possess if displaced adiabatically to a fiducial height in the atmosphere where $P_0=\hat{P}$, $\rho_0=\hat{\rho}$, and $s_0=0$. Like the anelastic approximation, the flow field can be expressed using streamfunctions when Equations~\eqnref{eqn:PI_continuity} and \eqnref{eqn:ref_potential_density} are valid,

\begin{equation}
    \label{eqn:PI_constraint}
    \grad\cdot \left(P_0^{1/\gamma} \bvec{u}\right)=0 \; ,
\end{equation}

\noindent We remind the reader that these two equations were derived using two assumptions: 1) the advective time scales are fast compared to diffusion times---i.e., isentropic motion, and 2) thermodynamic fluctuations are small compared to the background atmosphere.


\section{Internal Gravity Waves in a General Stratification}
\label{sec:IGW}

Consider a plane-parallel atmosphere with a gas pressure $P_0$ and mass density $\rho_0$ related through hydrostatic balance, $dP_0/dz=-g\rho_0$. Further, let the thermal structure of the atmosphere be general and specified by the vertical variation of the specific entropy density, $s_0$. We start with the linearized fluid equations for a fully-compressible ideal gas,

\begin{eqnarray}
    \label{eqn:Momentum}
    \rho_0\frac{\partial \bvec{u}}{\partial t} &=& -\nabla P_1 + \bvec{g} \rho_1 \; ,
\\
    \label{eqn:Entropy}
    \frac{\partial s_1}{\partial t} &=& -\bvec{u} \cdot \nabla s_0 \; ,
\\
    \label{eqn:Continuity}
    \frac{\partial \rho_1}{\partial t}  &=& - \nabla \cdot \left( \rho_0 \bvec{u}\right) \; .
\\
    \label{eqn:EOS}
    \frac{\rho_1}{\rho_0} &=& \frac{P_1}{\gamma P_0} - \frac{s_1}{c_p} \; . 
\end{eqnarray}

\noindent We have ignored rotation, magnetism, and all dissipative mechanisms, including viscosity, thermal conduction, and radiative transfer. The thermodynamic variables $s_1$, $\rho_1$, and $P_1$ are the Eulerian fluctuations of the specific entropy density, the mass density, and the gas pressure respectively. 

Since gravity provides the only preferred direction, internal gravity waves can be treated as a 2D phenomenon that propagates vertically and in a single horizontal direction. Let $\unitv{z}$ be the unit vector that is antiparallel to the constant gravitational acceleration, $\bvec{g} = - g\unitv{z}$. Further, let $\unitv{x}$ be the horizontal unit vector that is aligned with the wave's horizontal direction of propagation.  Finally, seek plane-wave solutions with the form

\begin{equation}
    \sim f(z) \, e^{ik_h x} \, e^{-i\omega t} \; ,
\end{equation}

\noindent where $k_h$ is the horizontal wavenumber, $\omega$ is the temporal frequency, and $f(z)$ is a vertical wave function.

The transformed set of equations can be manipulated to express the velocity and its divergence solely in terms of the Lagrangian pressure fluctuation, $\dP$.  The resulting equations are a coupled system of ODEs, 

\begin{eqnarray}
    \label{eqn:u}
    \rho_0 u &=& -\frac{\omega gk_h}{g^2k_h^2-\omega^4} \left(\frac{d}{dz} + \frac{\omega^2}{g}\right) \dP \; ,
\\
    \label{eqn:w}
    \rho_0 w &=& \frac{i\omega^3}{g^2k_h^2-\omega^4} \left(\frac{d}{dz} + \frac{gk_h^2}{\omega^2}\right) \dP \; ,
\\
    \label{eqn:div}
    \nabla \cdot \bvec{u} &=& \frac{i\omega}{\rho_0 c^2}\dP \; ,
\end{eqnarray}

\noindent with the vertical coordinate $z$ as the independent variable and $u$ and $w$ being the horizontal and vertical velocity components, $\bvec{u} = u \hat{\bvec{x}} + w \hat{\bvec{z}}$. The Lagrangian pressure fluctuation is related to the Eulerian pressure fluctuation and the vertical velocity,

\begin{eqnarray}
\nonumber
   \frac{\partial}{\partial t}\dP &\equiv& \frac{\partial P_1}{\partial t} + \bvec{u} \cdot \nabla P_0 \; ,
\\  \label{eqn:dP}
    \dP &=& P_1 + \frac{g \rho_0 w}{i\omega} \;.
\end{eqnarray}

\noindent The denominator of Equations~\eqnref{eqn:u} and \eqnref{eqn:w} is spatially constant and will appear later. Therefore for convenience we make the definition,
\begin{equation}
    \alpha \equiv g^2 k_h^2 - \omega^4 \; .
\end{equation}

Equations~\eqnref{eqn:u}--\eqnref{eqn:div} can be combined to produce a single stand-alone ODE with $\dP$ as the dependent variable,

\begin{equation}
    \label{eqn:ODE_dP}
    \left\{\frac{d^2}{dz^2} + \frac{1}{H}\frac{d}{dz} + {\frac{\omega^2}{c^2}} - k_h^2 \left(1-{\frac{N^2}{\omega^2}}\right)\right\}\dP = 0\;,
\end{equation}

\noindent where $N$ is the buoyancy frequency and $H$ is the density scale height,

\begin{eqnarray}
    N^2(z) &\equiv& g\left(\frac{1}{H} - \frac{g}{c^2}\right) = \frac{g}{c_p} \frac{ds_0}{dz} \; ,
\\
    \frac{1}{H(z)} &\equiv& -\frac{1}{\rho_0} \frac{d\rho_0}{dz} \; .
\end{eqnarray}

\noindent In Equation~(\ref{eqn:ODE_dP}), the term that involves the sound speed is responsible for the propagation of high-frequency acoustic waves and the term with the buoyancy frequency leads to internal gravity waves. As we will see in the following subsection, the first-derivative term ensures energy conservation for both varieties of wave.

Once one has solved for the Lagrangian pressure fluctuation by applying boundary conditions to Equation (\ref{eqn:ODE_dP}), the velocity components, $u$ and $w$, can be found directly through the use of Equations (\ref{eqn:u}) and  (\ref{eqn:w}). Subsequently, all of the thermodynamic fluctuations can then be derived through Equations~\eqnref{eqn:Entropy}, \eqnref{eqn:EOS}, and \eqnref{eqn:dP},

\begin{equation}
    \label{eqn:thermo_variables}
    \begin{split}
        P_1 &= \frac{k_h}{\omega} \rho_0 u \; , \quad s_1 = \frac{c_p N^2}{i\omega g} w\; ,
    \\
        \rho_1 &= \frac{\omega}{k_h c^2} \rho_0 u -  \frac{N^2}{i\omega g} \rho_0 w \; .
    \end{split}
\end{equation}

\noindent All of the thermodynamic fluctuations appear as linear combinations of the two velocity components.


\subsection{Energy Conservation and the First Derivative}
\label{subsec:first_derivative}

Here we demonstrate that any viable sound-proofing technique must produce an appropriate coefficient for the first-derivative term that appears in Equation~\eqnref{eqn:ODE_dP}. This term is crucial for energy conservation. To see this, consider the vertical energy flux for an acoustic-gravity wave, $F(z) = \left<w \, P_1\right>$, where angular brackets $<>$ indicate a temporal average over a wave period. Since, the second term on the right-hand side of Equation \eqnref{eqn:dP} is 90 degrees out of phase with the vertical velocity, in a time average the second term's contribution vanishes and the energy flux can be written just in terms of the Lagrangian pressure fluctuation,

\begin{equation}
    F(z) = \left<w \, \dP\right> = \frac{1}{4} \left(w \, \dP^* + w^* \, \dP\right)\; , 
\end{equation}

\noindent where the superscript asterisks denote complex conjugation. By employing Equation~(\ref{eqn:w}), one can demonstrate that this flux is inversely proportional to the mass density and proportional to the Wronskian of the Lagrangian pressure fluctuation and its complex conjugate,

\begin{equation}
    \label{eqn:Energy_Flux_full}
    F(z) = -\frac{i\omega^3}{4\alpha\rho_0} \left(\dP \frac{d \, \dP^*}{dz} - \dP^* \frac{d \, \dP}{dz} \right) \; .
\end{equation}

Abel's Identity tells us that to within an unknown multiplicative constant, $C$, the Wronskian depends only on the coefficient of the first derivative term in the ODE. For the ODE here, the necessary integration is trivial to perform,

\begin{equation}
    {\cal W}\left\{\dP,\dP^*\right\}(z) = C ~\exp\left(-\int \frac{dz}{H}\right) = C ~\rho_0\; .
\end{equation}

\noindent Hence, the energy flux is constant with height even though the coefficients of the ODE are vertically variable,

\begin{equation}
    F(z) = -\frac{i\omega^3 C}{4\alpha} = {\rm constant} \; .
\end{equation}

\noindent The constancy of the energy flux with height in the atmosphere is one way to characterize the conservation of energy by acoustic-gravity waves.

From this analysis, we can deduce that any approximation that incorrectly reproduces the first derivative term, may produce wave solutions with energy fluxes that vary with height. Consequently, such approximations will fail to conserve energy. For example, if the first derivative term is artificially set to zero, the flux will be inversely proportional to the mass density and $F(z)$ will spuriously increase with height.  This is the fundamental reason why \citet{Brown:2012} and \citet{Vasil:2013} found a lack of energy conservation when applying a variety of anelastic approximations to an isothermal atmosphere. Those approximations failed to correctly reproduce the first-derivative term of the ODE. Here we show that it is a general property for any stratification, not just an isothermal one.


\subsection{Local Dispersion Relation}
\label{subsec:dispersion_relation}

For a general stratification, the coefficients of the ODE~\eqnref{eqn:ODE_dP} are functions of height and the solutions will not be sinusoidal. However, by making a change of variable that converts the ODE into standard form (i.e., a Helmholtz equation that lacks a first-derivative term), a local dispersion relation can be generated which is appropriate in a WKB framework \citep[e.g.,][]{BO1999}. The required change of variable involves the square root of the mass density, $\dP = \left(\alpha \rho_0\right)^{1/2} \psi$. We include the constant $\alpha$ inside the square root purely for the sake of symmetry in later sections when we explore various sound-proofing techniques. Here, its inclusion is unnecessary and only introduces a multiplicative constant which factors out of the resulting ODE,

\begin{eqnarray}
    \label{eqn:ODE_psi}
    &&\frac{d^2\psi}{dz^2} + k_z^2\psi = 0\; ,
\\
    \label{eqn:dispersion}
    &&k_z^2(z) = \frac{\omega^2-\omega_c^2}{c^2} - k_h^2 \left(1-{\frac{N^2}{\omega^2}}\right) \; .
\end{eqnarray}

\noindent In the preceding equations, $k_z(z)$ is a local vertical wavenumber and $\omega_c(z)$ is the acoustic-cutoff frequency which depends on the stratification through the density scale height $H$,

\begin{equation}
    \frac{\omega_c^2}{c^2} \equiv \frac{1-2H^\prime}{4H^2} \; .
\end{equation}

\noindent We denote vertical derivatives of atmospheric quantities using a superscript prime, i.e., the vertical derivative of the density scale height is given by $H^\prime \equiv dH/dz$.

From the preceding analysis, we see that acoustic-gravity waves vary over two relevant vertical spatial scales: a local vertical wavelength and an envelope scale. The wavelength is given by the local dispersion relation~\eqnref{eqn:dispersion} and hence depends on the wave frequency as well as the characteristic frequencies of the atmosphere---i.e., the buoyancy frequency $N$, the acoustic cut-off frequency $\omega_c$, and the Lamb frequency $k_h c$.  The envelope scale is associated with vertical variation of the envelope function $\left(\alpha \rho_0\right)^{1/2}$ that appears in the change of variable above. This function provides a local amplitude of the wave function (in a WKB sense). Since the envelope function only depends on the mass density, the envelope scale is solely determined by the atmospheric stratification through the density scale height $H$. For later convenience, we choose to define the envelope scale $\Lambda$ as twice the scale length associated with the envelope function such that $\Lambda = H$,

\begin{equation}
  \Lambda^{-1} \equiv -\frac{2}{\left(\alpha \rho_0\right)^{1/2}} \frac{d \left(\alpha\rho_0\right)^{1/2}}{dz} =H^{-1} \;.
\end{equation}

\section{Internal Gravity Waves in the Low-Frequency Limit}
\label{sec:low-frequency_limit}

Our primary goal is to see how each wave variable scales with frequency and to therefore determine which terms are important in the fluid equations in the low-frequency limit. We start by non-dimensionalizing, using the reciprocal of the horizontal wavenumber $k_h^{-1}$ and the frequency of surface gravity waves $\sqrt{gk_h}$ for the characteristic length and frequency. We choose $c_p$ and $\hat{\rho}$ to be typical values of the entropy and mass density, respectively. Of particular importance is the non-dimensional wave frequency,

\begin{equation}
    \eps \equiv \frac{\omega}{\sqrt{g k_h}}
\end{equation}

\noindent which will serve as a small parameter in our low-frequency expansions.  Thus, when we speak of low frequencies we are considering frequencies that are small compared to those of surface gravity waves, $\omega^2 \ll gk_h$ or equivalently $\eps \ll 1$. This assumption will assure that the acoustic waves and the internal gravity waves decouple cleanly.

In combination, Equations~\eqnref{eqn:ODE_psi} and \eqnref{eqn:dispersion} indicate that the vertical wavelength of an internal gravity wave becomes very short as the frequency vanishes. To leading order in the frequency, the vertical wavenumber is determined by the ratio of the buoyancy frequency to the wave frequency,

\begin{equation}
    \label{eqn:ODE_order0}
    k_z^2 \approx  k_h^2 \frac{N^2}{\omega^2} \; .
\end{equation}

\noindent Hence, in the low-frequency limit the vertical wavelength becomes a {\sl short} spatial scale, whereas the envelope or atmospheric scale remains {\sl long}. This scale separation dictates that we must define a non-dimensional height $\zeta$ that appropriately rescales the vertical derivatives in the fluid equations to respect the short scale,

\begin{equation}
   \frac{d}{dz} \equiv \frac{k_h}{\eps} \frac{d}{d\zeta} \; .
\end{equation}
    
If we denote the non-dimensional forms of the wave variables and atmospheric profiles using a tilde, the wave equation~(\ref{eqn:ODE_dP}) becomes,

\begin{equation}
    \label{eqn:ODE_nondim}
    \left\{\frac{d^2}{d\zeta^2} + \frac{\eps}{\tilde{H}}\frac{d}{d\zeta} + \left[\tilde{N}^2 - \eps^2 + \frac{\eps^4}{\tilde{c}^2} \right]\right\}\delta \tilde{P} = 0\; ,
\end{equation}

\noindent where the non-dimensional atmospheric profiles are given by

\begin{equation}
    \tilde{H} = k_h H,   \qquad \tilde{c}^2 = \left(\frac{k_h}{g}\right) c^2, \qquad \tilde{N}^2 = \frac{N^2}{gk_h}
\end{equation}

\noindent and the non-dimensional Lagrangian pressure fluctuation is defined as follows:

\begin{equation}
    \delta\tilde{P} = \frac{k_h}{g\hat{\rho}} \dP \; .
\end{equation} 

Similarly, the non-dimensional form for the local dispersion relation is given by

\begin{equation}
\label{eqn:kz}
    \frac{k_z^2(z)}{k_h^2} = \eps^{-2} \tilde{N}^2  - \left(1 + \tilde{k}_c^2\right) + \frac{\eps^2}{\tilde{c}^2}  \; ,
\end{equation}

\noindent where $\tilde{k}_c$ is a nondimensional wavenumber that is the ratio of the acoustic cutoff frequency to the Lamb frequency,

\begin{equation}
    \tilde{k}_c^2 \equiv \frac{\omega_c^2}{k_h^2 c^2} = \frac{1 - 2H^\prime}{4\tilde{H}^2} \; .
\end{equation}.

\noindent As expected, the leading order behavior of the local vertical wavenumber in Equation~\eqnref{eqn:kz} demonstrates that the vertical wavelength becomes very short in the low-frequency limit, $k_z^2 / k_h^2 \sim \eps^{-2} \tilde{N}^2$. Modifications to the vertical wavenumber arising from a finite frequency first appear at order unity, ${\cal O}(1)$, whereas the term in the dispersion relation responsible for the propagation of high-frequency acoustic waves appears at ${\cal O}(\eps^2)$.


\subsection{Frequency Dependence of the Other Wave Variables}
\label{subsec:other_variables}

The non-dimensional forms of the other fluid variables can be generated through Equations~\eqnref{eqn:u}, \eqnref{eqn:w}, and \eqnref{eqn:thermo_variables}
and are related to the Lagrangian pressure fluctuation through differential operators,

\onecolumngrid
\begin{alignat}{3}
    \label{eqn:nondim_u}
    \tilde{u} = \left(\frac{k_h}{g}\right)^{1/2} u &= - \frac{\tilde{\rho}_0^{-1}}{1-\eps^4}\left(\frac{d}{d\zeta} + \eps^3\right) \delta\tilde{P}  &\sim {\cal O}(1) \; ,
\\
    \label{eqn:nondim_w}
    \tilde{w} = \left(\frac{k_h}{g}\right)^{1/2} w &= \frac{i\eps ~\tilde{\rho}_0^{-1}}{1-\eps^4}\left(1 + \eps \frac{d}{d\zeta}\right) \delta\tilde{P} &\sim {\cal O}(\eps) \; ,
\\
    \tilde{P}_1 = \left(\frac{k_h}{g\hat{\rho}}\right) P_1 &= - \frac{\eps}{1-\eps^4}\left(\frac{d}{d\zeta} + \eps^3\right) \delta\tilde{P}  &\sim {\cal O}(\eps) \; ,
\\
    \tilde{s}_1 = \frac{s_1}{c_p} &= \frac{\tilde{N}^2\tilde{\rho}_0^{-1}}{1-\eps^4}\left(1 + \eps \frac{d}{d\zeta}\right) \delta\tilde{P}   &\sim {\cal O}(1) \; ,
\\
    \label{eqn:nondim_density}
    \tilde{\rho}_1 = \frac{\rho_1}{\hat{\rho}} &= -\frac{1}{1-\eps^4}\left(\tilde{N}^2 + \frac{\eps}{\tilde{H}}\frac{d}{d\zeta} + \frac{\eps^4}{\tilde{c}^2} \right) \delta\tilde{P}~~&\sim{\cal O}(1)\; ,
\end{alignat}
\twocolumngrid

\noindent where $\tilde{\rho}_0 = \rho_0 / \hat{\rho}$ is the non-dimensional atmospheric density.

We can immediately see that internal gravity waves possess motions that are nearly horizontal for low frequencies. The vertical velocity component $w$ is small by a factor of $\eps$. Furthermore, while the Lagrangian pressure fluctuation remains order unity in size, $\dP \sim {\cal O}(1)$, the Eulerian pressure fluctuation becomes small, $P_1 \sim {\cal O}(\eps$). Both the entropy and density fluctuations remain order unity. The fact that the Eulerian pressure fluctuation vanishes in the limit of low frequency is consistent with the pseudo-incompressible approximation and ensures that the internal gravity waves and acoustic waves decouple in that limit. However, since the mass density fluctuation does not vanish, these limits further suggest that this decoupling is {\bf not} accomplished through the anelastic limit. We explore this result fully in the next subsection.

In order to make obvious the relative magnitude of terms in subsequent equations, we define alternate dimensionless variables for the vertical velocity and Eulerian pressure fluctuation,

\begin{equation}
    \tilde{w} \equiv \eps \tilde{W} \; ,   \qquad \tilde{P}_1 \equiv \eps \tilde{\Theta} \; .
\end{equation}

\noindent Both $\tilde{W}$ and $\tilde{\Theta}$ are order unity because the prefactors in their definitions absorb the leading-order behavior as the frequency becomes small.


\subsection{Low-Frequency Limit of the Continuity Equation}
\label{subsec:continuity_equation}

Consider the dimensional form of the continuity equation~(\ref{eqn:Continuity}), where the equation of state~(\ref{eqn:EOS}) is used to replace the density fluctuation

\begin{equation}
    \label{eqn:continuity_EOS}
    i\omega \frac{\rho_1}{\rho_0} = i\omega \left(\frac{P_1}{\rho_0 c^2} - \frac{s_1}{c_p}\right) = ik_h u + \frac{dw}{dz} - \frac{w}{H} \; .
\end{equation}

\noindent In order to sound proof the equation set, we need to eliminate the term involving the Eulerian pressure fluctuation. This term is responsible for producing the pressure fluctuations that generate the restoring force for acoustic oscillations.

The anelastic approximation does indeed eliminate this pressure term, but it is overkill and removes the entire left-hand side of the continuity equation above.  In particular, the term involving the entropy fluctuation is also thrown away. For low-frequency internal gravity waves, this is inconsistent.  If the continuity equation is non-dimensionalized it becomes obvious that the entropy term is the same order as other terms that are retained by the anelastic approximation,

\begin{equation}
    \label{eqn:colored_continuity}
    i\eps^2 \frac{\tilde{\Theta}}{\tilde{\rho}_0 \tilde{c}^2} - i\eps \tilde{s}_1 = \left[i \tilde{u} + \frac{d\tilde{W}}{d\zeta}\right] - \eps \frac{\tilde{W}}{\tilde{H}} \; .
\end{equation}

\noindent The leading-order behavior consists of the two order-unity terms that appear in square brackets on the right-hand side of Equation~\eqnref{eqn:colored_continuity}. The first correction for nonzero frequency is comprised of the two first-order terms, ${\cal O}(\eps)$; one of these is the foresaid entropy term. The term involving the Eulerian pressure fluctuation is second order, ${\cal O}(\eps^2)$.

The lowest-order self-consistent approximation that one could make would be to keep just the leading-order terms, resulting in an assumption of incompressibility, $\nabla \cdot \bvec{u} \approx 0$. The next self-consistent approximation would be the retention of all zero-order and first-order terms. As we will show next, this approximation is equivalent to the pseudo-incompressible condition.

We demonstrate pseudo-incompressibility by using the energy equation~\eqnref{eqn:Entropy} to replace the entropy fluctuation in Equation~\eqnref{eqn:continuity_EOS} with the vertical velocity and then combining the two first-order terms using the definition of the buoyancy frequency, $N^2 = g/H - g^2/c^2$,

\begin{equation}
    i\omega \frac{P_1}{\rho_0 c^2} = \left[ik_h u + \frac{dw}{dz}\right] - \frac{gw}{c^2} \; .
    \label{eqn:continuity_no_s}
\end{equation}

\noindent The last term on the right-hand side is equal to the vertical velocity divided by the scale height for the potential density, i.e., the density scale height for an adiabatic density stratification,

\begin{equation}
    \frac{1}{H_*} \equiv - \frac{1}{\rho_{*0}} \frac{d\rho_{*0}}{dz} = \frac{g}{c^2} \; .
\end{equation}

\noindent Hence, the terms on the right-hand side of Equation~\eqnref{eqn:continuity_no_s} can be cleanly combined,

\begin{equation}
    \label{eqn:pseudo-incompressible}
    \nabla \cdot \left(\rho_{*0} \bvec{u}\right) = i\omega \frac{\rho_{*0} P_1}{\rho_0 c^2} \sim {\cal O}(\eps^2) \; .
\end{equation}

\noindent A self-consistent low-frequency approximation is to discard all second-order terms, leading to the pseudo-incompressible approximation, $\grad\cdot\left(\rho_{*0} \bvec{u}\right)=0$.

\subsection{Low-Frequency Limit of the Momentum Equation}
\label{subsec:vertical_momentum}

When transformed into the spectral representation, the vertical component of the momentum equation~\eqnref{eqn:Momentum} is given by

\begin{equation}
    \label{eqn:vertical_momentum}
    -i\omega \rho_0 w = -\frac{dP_1}{dz} - g\rho_1 \; ,
\end{equation}

\noindent and non-dimensionalization of this equation yields,

\begin{equation}
    -i \eps^2 \tilde{\rho}_0 \tilde{W} = -\frac{d\tilde{\Theta}}{d\zeta} - \tilde{\rho}_1 \; .  
\end{equation}

\noindent It is now obvious from the preceding equation that the inertial term on the left-hand side becomes the smallest term in the low-frequency limit; it is a second-order correction. The right-hand side consists solely of terms that are zero order in the dimensionless frequency. Hence, to first order, the balance is simply the hydrostatic relation between the perturbed pressure and the perturbed density,

\begin{equation}
    -\frac{dP_1}{dz} - g\rho_1 \approx 0 \; .
\end{equation}

The pseudo-incompressible and fiducial anelastic approximations both leave the vertical momentum equation unmodified.  But, the LBR formulation of the anelastic approximation drops a term whose removal is formally valid only in an adiabatic (or near-adiabatic) stratification. The vertical momentum equation~\eqnref{eqn:vertical_momentum} can be rewritten in the following manner

\begin{equation}
    \label{eqn:LBR_momentum}
    -i \omega w = -\frac{d}{dz} \left(\frac{P_1}{\rho_0}\right) + \frac{g s_1}{c_p} + \frac{N^2}{g} \frac{P_1}{\rho_0} \; .  
\end{equation}

\noindent either by linearizing Equation~\eqnref{eqn:anelastic_momentum} or by dividing the vertical momentum equation~\eqnref{eqn:vertical_momentum} by the mass density and pulling the density into the gradient operator that appears in the pressure force by use of the chain rule. The LBR formulation of the anelastic approximation removes the term involving the buoyancy frequency, even in stable stratifications where the buoyancy frequency is not small. We demonstrate that this approximation is inconsistent with low-frequency gravity waves by considering the nondimensional form of Equation~\eqnref{eqn:LBR_momentum},

\begin{equation}
    -i \eps^2 \tilde{W} = - \frac{d}{d\zeta} \left(\frac{\tilde{\Theta}}{\tilde{\rho}_0}\right) + \tilde{s}_1 + \eps\tilde{N}^2 \frac{\tilde{\Theta}}{\tilde{\rho}_0} \; . 
\end{equation}

\noindent The LBR approximation inconsistently ignores the first-order term while retaining the inertial term (which is second-order).


\section{The Integrity of Three Sound-Proofing Treatments}
\label{sec:sound-proofing}

In this section we examine the success or failure of a variety of sound-proofing methods in reproducing the appropriate behavior of low-frequency internal gravity waves. We have already discussed how all anelastic formulations inconsistently reject terms in the continuity equation and how the LBR anelastic formulation is further inconsistent with its treatment of the vertical momentum equation. Here we will examine how these inconsistencies propagate and produce errors in the dispersion relation and wave functions. To ease comparison, here we provide the local dispersion relation for a fully compressible fluid in both its dimensional and nondimensional forms---i.e., Equations~\eqnref{eqn:dispersion} and \eqnref{eqn:kz},

\begin{eqnarray}
    k_z^2(z) &=& k_h^2 \left({\frac{N^2}{\omega^2}}-1\right) - \frac{\omega_c^2}{c^2} + \frac{\omega^2}{c^2} \; ,
\\
    \frac{k_z^2(z)}{k_h^2} &=& \eps^{-2} \tilde{N}^2  - \left(1 + \tilde{k}_c^2\right) + \frac{\eps^2}{\tilde{c}^2}  \; .
\end{eqnarray}

\noindent Further, in Table~\ref{tab:table1}, we summarize the function $\alpha(z)$, the local wavenumber $k_z$, and the envelope scale $\Lambda$ in the low-frequency limit for a fully compressible fluid and for all three sound-proofing treatments.  We retain terms only up to first-order in the dimensionless frequency $\eps$.

\tableone

\subsection{Pseudo-incompressible Approximation}
\label{subsec:PI}

Since the pseudo-incompressible approximation is self-consistent in its treatment of the continuity equation and correct to first order in the frequency, we expect that this approximation should produce low-frequency internal gravity waves that are correct to first order in the dispersion relation and in the wave functions. To demonstrate that this expectation is true we rederive the wave equation for internal gravity waves but with the continuity equation~\eqnref{eqn:Continuity} replaced by the pseudo-incompressible condition, $\grad \cdot \left(\rho_{*0} \bvec{u}\right) = 0$. We simply present the result,

\begin{equation}
    \label{eqn:PI_dim}
    \begin{split}
        &\left\{ \frac{d^2}{dz^2} + \left(\frac{1}{H} + \frac{\omega^2}{g}\theta_{\rm PI}\right)\frac{d}{dz} - k_h^2 \left(1 - \frac{N^2}{\omega^2}\right)\right.
        \\
        &\qquad\qquad\qquad\qquad \left.+ \left[\frac{N^2}{c^2} + \frac{\omega^4}{g^2}\theta_{\rm PI}\right] \right\} \dP = 0 \; .
    \end{split}
\end{equation}

\noindent In this expression, $\theta_{\rm PI}(z)$ is a dimensionless function that depends on the temporal frequency $\omega$, the horizontal wavenumber $k_h$, and the potential density $\rho_{*0}$ through the following definitions

\begin{eqnarray}
    \label{eqn:PI_alpha}
    \alpha_{\rm PI}(z) &\equiv& g^2k_h^2 - \omega^4 + \omega^2 \frac{g}{H_*} \; ,
\\
    \label{eqn:PI_theta}
    \theta_{\rm PI}(z) &\equiv& -\frac{g}{\omega^2} \frac{\alpha_{\rm PI}^\prime}{\alpha_{\rm PI}} = \frac{g^2}{\alpha_{\rm PI}} \, \frac{H_*^\prime}{H_*^2} \; .
\end{eqnarray}

\noindent Compared to the fully-compressible equations, the quantity $\alpha_{\rm PI}$ has been augmented by $\omega^2 g/H_*$, and is therefore no longer a constant function of height.

A direct comparison of Equation~\eqnref{eqn:PI_dim} with the wave equation for a fully compressible fluid~\eqnref{eqn:ODE_dP} reveals that there are three spurious terms: both of the terms involving $\theta_{\rm PI}$, as well as the term $(N^2/c^2)\delta P$. To demonstrate that all of these spurious terms are small in magnitude and can be safely ignored in the low-frequency limit, we nondimensionalize Equation~\eqnref{eqn:PI_dim},

\begin{equation}
    \label{eqn:PI_nondim}
    \begin{split}
        &\left\{ \frac{d^2}{d\zeta^2} + \left(\frac{\eps}{\tilde{H}} + \eps^3\theta_{\rm PI}\right)\frac{d}{d\zeta} + \tilde{N}^2  -\eps^2\right.
    \\
        &\qquad\qquad\qquad\qquad  + \left.\left[\eps^2\frac{\tilde{N}^2}{\tilde{c}^2}  + \eps^6 \theta_{\rm PI}\right] \right\} \delta\tilde{P} = 0 \; ,
    \end{split}
\end{equation}

\noindent and we recognize that the function $\theta_{\rm PI}$ is an order-unity quantity for low frequencies,

\begin{equation}
    \theta_{\rm PI} = \frac{1}{1-\eps^4 + \eps^2\tilde{H}_*^{-1}} \, \frac{H_*^\prime}{\tilde{H}_*^2} \quad \sim {\cal O}(1) \; .
\end{equation}

\noindent Thus, all of the spurious terms are second-order or higher in the dimensionless frequency $\eps$ and the Lagrangian pressure fluctuation that is generated by Equation~\eqnref{eqn:PI_dim} is correct to first-order.

Based on this result we should expect the local dispersion relation to also be correct to first order and this is indeed the case. The transformation that converts the ODE into a Helmholtz equation has the same form as we found for the fully-compressible equations,

\begin{equation}
    \delta P = \left(\alpha_{\rm PI} \, \rho_0 \right)^{1/2}  \psi \; , 
\end{equation}

\noindent but now the function $\alpha = \alpha_{\rm PI}(z)$ varies with height. This change of variable leads to the following local dispersion relation,

\begin{equation}
    \begin{split}
        &k_z^2(z) = k_h^2 \left(\frac{N^2}{\omega^2}-1\right) -\frac{\omega_c^2}{c^2} 
    \\
        &\, + \left[\frac{N^2}{c^2} - \frac{\omega^2}{2g}\left(\theta_{\rm PI}^\prime + \frac{\theta_{\rm PI}}{H}\right)
        + \frac{\omega^4}{g^2}\left(\theta_{\rm PI}-\frac{\theta_{\rm PI}^2}{4}\right)\right] \, ,
    \end{split}
\end{equation}

\noindent with a nondimensional form given by

\begin{equation}
    \begin{split}
        &\frac{k_z^2(z)}{k_h^2} = \eps^{-2} \tilde{N}^2  -  \left(1 + \tilde{k}_c^2\right)
    \\
        & \; + \left[\frac{\tilde{N}^2}{\tilde{c}^2} - \frac{\eps^2}{2} \left(\frac{\theta_{\rm PI}^\prime}{k_h} + \frac{\theta_{\rm PI}}{\tilde{H}}\right)
         + \eps^4\left(\theta_{\rm PI} - \frac{\theta_{\rm PI}^2}{4}\right)\right]  \; .
    \end{split}
\end{equation}

\noindent All of the terms contained by the error term, $E_{\rm PI}$, in the preceding equations are spurious and do not appear in the local dispersion relation for a fully compressible fluid.  However, all spurious terms appear as a correction that is smaller than the leading order behavior by a factor of $\eps^2$ or smaller. Hence, the pseudo-incompressible approximation leads to a local dispersion relation that is correct to first order.

Finally, the envelope scale can be read directly from the coefficient of the first-derivative term in the ODE, $\Lambda^{-1} = H^{-1} + \omega^2\theta_{\rm PI}/g$. To first order in the frequency, the envelope scale is simply the density scale height.

\subsection{Fiducial Anelastic}
\label{subsec:ANS}

For the fiducial anelastic approximation, where the only modification to the fully-compressible fluid equations is made to the continuity equation, the resulting ODE for the Lagrangian pressure fluctuation is as follows,

\begin{equation}
    \label{eqn:FA_dim}
    \begin{split}
        &\left\{ \frac{d^2}{dz^2} + \left(\frac{1}{H_*} + \frac{\omega^2}{g}\theta_{\rm FA}\right) \frac{d}{dz} - k_h^2 \left(1 - \frac{N^2}{\omega^2}\right)\right.
    \\
        &\qquad\qquad\; \left.+ \left[ \left(\frac{\omega^2}{c^2} + k_h^2\right)\theta_{\rm FA} - \frac{H_*^\prime}{H_*^2} \right] \right\} \dP = 0 \; ,
    \end{split}
\end{equation}

\noindent where the $\alpha$ and $\theta$ functions take on subtly but crucially different forms,

\begin{eqnarray}
    \label{eqn:FA_alpha}
    \alpha_{\rm FA}(z) &\equiv& g^2 k_h^2 - \omega^4 + \omega^2 \frac{g}{H} \; ,
\\
    \theta_{\rm FA}(z) &\equiv& -\frac{g}{\omega^2} \frac{\alpha_{\rm FA}^\prime}{\alpha_{\rm FA}} = \frac{g^2}{\alpha_{\rm FA}} \, \frac{H^\prime}{H^2} \; .
\end{eqnarray}

\noindent Here, $\alpha_{\rm FA}$ and $\theta_{\rm FA}$ differ from the pseudo-incompressible case, Equations~\eqnref{eqn:PI_alpha} and \eqnref{eqn:PI_theta}, by the appearance of $H$ instead of $H_*$.

A direct comparison of Equation~\eqnref{eqn:FA_dim} with the ODE~\eqnref{eqn:ODE_dP} appropriate for a fully compressible fluid illustrates that fiducial anelastic generates a variety of spurious and incorrect terms. Specifically, the terms in the square brackets are spurious and the entire coefficient of the first-derivative term is incorrect. To ascertain the magnitude of these mistakes, we nondimensionalize,

\begin{equation}
    \label{eqn:FA_nondim}
    \begin{split}
        &\left\{ \frac{d^2}{d\zeta^2} + \left(\frac{\eps}{\tilde{H}_*} + \eps^3\theta_{\rm FA} \right)\frac{d}{d\zeta} + \tilde{N}^2 - \eps^2\right.
    \\
        &\qquad\qquad\; + \left.\eps^2\left(\theta_{\rm FA}-\frac{H_*^\prime}{\tilde{H}_*^2} \right) + \eps^4\frac{\theta_{\rm FA}}{\tilde{c}^2} \right\} \delta\tilde{P} = 0 \; ,
    \end{split}
\end{equation}

\noindent Fiducial anelastic performs rather poorly in reproducing the behavior of low-frequency internal gravity waves. The ODE is correct only to leading order in $\eps$ with inconsistencies appearing at first-order in the coefficient of the first derivative. The first term in this coefficient contains the reciprocal of the scale height of the potential density, where it should instead possess the reciprocal of the density scale height---see Equation~\eqnref{eqn:ODE_nondim}.

Interestingly, conversion of the ODE to standard form---via the change of variable $\delta P = \left(\alpha_{\rm FA} \, \rho_*\right)^{1/2} \psi$---results in a local dispersion relation that is correct to first order,

\begin{equation}
    \begin{split}
    \label{eqn:fiducial_disp}
        &k_z^2(z) = k_h^2 \left(\frac{N^2}{\omega^2} - 1\right) - \frac{1+2H_*^\prime}{4H_*^2}
    \\
        &\qquad\; + \left[k_h^2\theta_{\rm FA} - \frac{\omega^2}{2g} \left(\theta_{\rm FA}^\prime - \frac{\theta_{\rm FA}}{H_*}\right) - \frac{\omega^4}{4g^2}\theta_{\rm FA}^2\right] \; ,
    \end{split}
\end{equation}

\noindent or

\begin{equation}
    \begin{split}
        &\frac{k_z^2(z)}{k_h^2} = \eps^{-2}\tilde{N}^2 - \left(1 + \frac{1+2H_*^\prime}{4\tilde{H}_*^2}\right)
    \\
        &\qquad\; + \left[\theta_{\rm FA} + \frac{\eps^2}{2} \left(k_h^{-1}\theta_{\rm FA}^\prime - \frac{\theta_{\rm FA}}{\tilde{H}_*}\right) - \frac{\eps^4}{4}\theta_{\rm FA}^2\right] \; .
    \end{split}
\end{equation}

\noindent In addition to all of the spurious terms that appear in the square brackets, the acoustic cut-off frequency is incorrect,

\begin{equation}
    \frac{1+2H_*^\prime}{4\tilde{H}_*^2} \neq \tilde{k}_c^2 = \frac{1-2H^\prime}{4\tilde{H}^2} \; .
\end{equation}

\noindent For ease of comparison, in Table~\ref{tab:table1} we have reworked the right-hand side of Equation~\eqnref{eqn:fiducial_disp} to extract the correct form of the acoustic cutoff frequency. Despite these issues, the errors all appear at second order or higher in the dimensionless frequency $\eps$, meaning that the erroneous terms divided by the leading order behavior are small by a factor of $\eps^2$. The fact that the ODE itself is incorrect at first order manifests in the envelope function, $\left(\alpha_{\rm FA} \, \rho_*\right)^{1/2}$, which is wrong at all orders. As we will see in a subsequent section this results in first-order errors in the wave functions even though the dispersion relation is correct to first order.

\subsection{LBR Anelastic}
\label{subsec:LBR}

In the framework of the LBR anelastic approximation, in addition to the anelastic treatment of the continuity equation, i.e., $\grad \cdot\left(\rho_0 \bvec{u}\right) \approx 0$, a term in the vertical momentum equation is removed. When these two modification to the fluid equations are adopted, the resulting ODE that describes internal gravity waves becomes,

\begin{equation}
    \label{eqn:LBR_dim}
    \begin{split}
        &\left\{ \frac{d^2}{dz^2} + \left(\frac{1}{H} + \frac{\omega^2}{g}\theta_{\rm LBR}\right)\frac{d}{dz} - k_h^2 \left(1 - \frac{N^2}{\omega^2}\right)\right.
    \\
        &\qquad\quad + \left.\left[\left(k_h^2 + \frac{\omega^2}{gH} \right) \theta_{\rm LBR} - \frac{H^\prime}{H^2}\right] \right\} \dP = 0 \; ,
    \end{split}
\end{equation}

\noindent where $\alpha$ and $\theta$ are now

\begin{eqnarray}
    \label{eqn:LBR_alpha}
    \alpha_{\rm LBR}(z) &\equiv& g^2k_h^2 - \omega^4 +\omega^2\left(N^2+\frac{g}{H}\right)\; ,
\\
    \theta_{\rm LBR}(z) &\equiv& \frac{g^2}{\alpha_{\rm LBR}}\left(\frac{H^\prime}{H^2} - \frac{1}{g}\frac{dN^2}{dz}\right) \; .
\end{eqnarray}

The non-dimensional form of the ODE becomes

\begin{equation}
    \begin{split}
        &\left\{ \frac{d^2}{d\zeta^2} + \left(\frac{\eps}{\tilde{H}} + \eps^3\theta_{\rm LBR} \right)\frac{d}{d\zeta} + \tilde{N}^2 - \eps^2 \right.
    \\
        &\qquad\; + \left.\left[\eps^2 \left(\theta_{\rm LBR} - \frac{H^\prime}{\tilde{H}^2}\right) + \eps^4\frac{\theta_{\rm LBR}}{\tilde{H}}\right] \right\} \delta\tilde{P} = 0 \; .
    \end{split}
\end{equation}

\noindent Despite the inconsistent treatment of the vertical momentum equation, the LBR form of the anelastic approximation generates an ODE that is correct to first order in $\eps$. The spurious terms that appear in the square brackets are second order or higher and the coefficient of the first derivative is correct to first order. As expected the local dispersion relation---once again achieved by the change of variable $\delta P = \left(\alpha_{\rm LBR} \, \rho_0 \right)^{1/2} \psi$, is correct to first order,

\begin{equation}
    \begin{split}
        &k_z^2(z) = k_h^2\left(\frac{N^2}{\omega^2} -1\right) - \frac{\omega_c^2}{c^2} - \frac{H^\prime}{H^2}
    \\
        &\quad +  k_h^2 \theta_{\rm LBR} - \frac{\omega^2}{2g}\left( \theta_{\rm LBR}^\prime - \frac{\theta_{\rm LBR}}{H}\right) -\frac{\omega^4}{4g^2}\theta_{\rm LBR}^2 \; ,
    \end{split}
\end{equation}

\noindent and

\begin{equation}
    \begin{split}
        &\frac{k_z^2(z)}{k_h^2} = \eps^{-2}\tilde{N}^2 - \left(1 + \tilde{k}_c^2 + \frac{H^\prime}{\tilde{H}^2}\right)
    \\
        &\quad\; + \theta_{\rm LBR} - \frac{\eps^2}{2}\left(k_h^{-1}\theta_{\rm LBR}^\prime - \frac{\theta_{\rm LBR}}{\tilde{H}}\right)-\frac{\eps^4}{4} \theta_{\rm LBR}^2 \; .
    \end{split}
\end{equation}
\vspace{1cm}

\subsection{Comparison of the Vertical Wavelengths}
\label{subsec:wavelengths}

In the previous subsections we demonstrated that the three approximations generate errors to the vertical wavelength of internal gravity waves that are second order in the dimensionless frequency $\eps$. Hence, if the only test of fidelity was to reproduce the local dispersion relation, all of the sound-proofing treatments would fair equally well. This is born out by a comparison of the vertical wavenumber that is achieved in an isothermal atmosphere by each treatment. This type of atmosphere is one of the most lenient of all potential atmospheres because all of the characteristic frequencies, i.e., $N$, $\omega_c$, and $k_h c$, become constant functions of height, as do the scale heights $H$ and $H_*$. As a consequence, the vertical wavenumber $k_z$ becomes a constant and the quantity $\theta$ vanishes identically for all approximations. When $\theta$ is zero, many of the spurious terms disappear from the local dispersion relations.

\isothermal

Figure~\ref{fig:isothermal} shows the performance of the three approximations in an isothermal atmosphere. The left-most panel illustrates the isocontours of the vertical wavenumber achieved in a fully-compressible fluid as a function of horizontal wavenumber $k_h$ and temporal frequency $\omega$. The remaining three panels provide the same isocontours for the sound-proofing treatment indicated at the top of the panel. The solid black contours in each panel are for the fully-compressible fluid, while the dashed red curves show the same contours under the relevant approximation. The value of each contour is marked in the left-most panel. In each panel, four isocontours of the nondimensional frequency $\eps = \omega/\sqrt{gk_h}$ are overlayed for reference and appear as dotted orange curves. To see how well an approximation reproduces the correct behavior, one should compare the red and black curves within a panel.  We would expect that the differences should be small for low values of the nondimensional frequency, i.e., in the lower-right portion of the diagram, and large for high values of $\eps$ (upper left).
From the four panels, it is clear that all three approximations reproduce the vertical wavenumber well as long as the dimensionless frequency is small, i.e., $\eps \lesssim 0.3$.

\subsection{Comparison of Wave Cavity Boundaries}
\label{subsec:wave_cavity}

Since an isothermal atmosphere is so special (because many of the spurious terms in the dispersion relation vanish), it is wise to examine the behavior of the local dispersion relation in a more complicated atmosphere. We have chosen to examine the vertical wavenumber in a polytropically stratified atmosphere.  Such atmospheres have thermodynamic profiles that are power laws in the depth,

\begin{eqnarray}
    \rho_0(z) &=& A (-z)^m \;,  \quad P_0(z) = \frac{A g}{m+1} (-z)^{m+1} \; ,
\\ 
    H(z) &=& \frac{(-z)}{m} \;, \quad  N^2(z) = \frac{m(\gamma-1)-1}{\gamma}\frac{g}{(-z)} \; , 
\\
    H_*(z) &=& \frac{m+1}{m}(-z) \;, \quad c^2(z) = \frac{\gamma g}{m+1} (-z) \; .
\end{eqnarray}

\noindent In the expressions above, $A$ is an arbitrary constant and $m$ is the polytropic index.  Polytropes can be stably or unstably stratified depending on the values of the adiabatic index $\gamma$ and the polytropic index $m$; if $m>(\gamma-1)^{-1}$, the atmosphere is stable to convective overturning.

\polytrope

A convenient feature of a polytropic atmosphere is that it is self-similar, lacking an intrinsic spatial scale \citep[see][]{Hindman:2022}.  Therefore, the local dispersion relation becomes independent of the horizontal wavenumber if we express the frequency in terms of our nondimensional frequency $\eps =\omega/\sqrt{gk_h}$ and we write all of the atmospheric profiles using a nondimensional depth $-k_h z$. Because of this property, we can generate a single dispersion diagram that illustrates the vertical wavenumber as a function of dimensionless depth and frequency that is valid for all horizontal wavenumbers.  Figure~\ref{fig:polytrope} presents the resulting dispersion diagram for each treatment of the fluid equations for a polytropic atmosphere with a polytropic index of $m=3$ (which is stably stratified for an adiabatic index of $\gamma=5/3$). The left-most panel is for a fully-compressible fluid and the right three panels are for the three sound-proofing formalisms. The blue region in each diagram corresponds to those depths in the atmosphere where a wave of the given frequency is vertically evanescent and the black and red contours have the same meaning as in Figure~\ref{fig:isothermal}. The upper panels show a range of dimensionless frequency that is wide enough to contain both the branch of low-frequency internal gravity waves and the branch of high-frequency acoustic waves (if present). The lower panels show a zoom-in view at low-frequencies that focuses on the gravity waves.  Note, at a given frequency, there are two turning points where the local vertical wavenumber vanishes. Hence, the internal gravity waves are vertically trapped in a wave cavity for $g$ modes. The turning points are indicated by the thick curves. Similarly, in the fully-compressible fluid, the acoustic waves are trapped in a $p$ mode cavity.

The pseudo-incompressible condition does minimal damage to the $g$-mode cavity, see Figure~\ref{fig:polytrope}b.  The boundaries move only slightly even for the highest-frequency waves. Further, the vertical wavenumbers within the cavity are weakly affected even for frequencies that are large enough that we might suspect that the low-frequency limit is invalid. The anelastic models fare poorly, however. The fiducial anelastic approximation does a horrendous job of reproducing the wave cavity boundaries.  In fact, there appears to be a residual of the acoustic cavity that is highly distorted and appears at frequencies halfway between the acoustic and gravity wave branches. While, the LBR form of the anelastic approximation does not have spurious wave cavities at high frequency, it fails to reproduce the boundaries of the $g$-mode cavity with fidelity.  The highest-frequency gravity waves that are vertically propagating have frequencies that are too high by a factor of about one-third. Further, errors in the vertical wavenumber become noticably large for relatively low values of the dimensionless frequency, $\eps > 0.1$.

\subsection{Errors in the Wave Functions}
\label{subsec:wave functions}

In sections~\ref{subsec:PI}--\ref{subsec:LBR}, we found that the pseudo-incompressible approximation and the LBR formulation of the anelastic approximation both introduced errors in the Lagrangian pressure fluctuation that appeared at second order. The fiducial anelastic approximation produced errors at first-order.  So at first glance the LBR approximation seems to fare well.  However, as we shall soon see, when we consider other wave variables, such as the fluid velocity components, the pseudo-incompressible approximation becomes the clear winner.

In the same manner that one derives equations~\eqnref{eqn:u} and \eqnref{eqn:w} for a fully compressible fluid, similar equations can be derived for each of the approximations. When pseudo-incompressibility is adopted, one obtains the following:

\begin{eqnarray}
    \label{eqn:u_PI}
    \rho_0 u &=& -\frac{\omega gk_h}{\alpha_{\rm PI}} \left(\frac{d}{dz} + \frac{\omega^2}{g}\right) \dP \; ,
\\
    \label{eqn:w_PI}
    \rho_0 w &=& \frac{i\omega^3}{\alpha_{\rm PI}} \left(\frac{d}{dz} + \frac{gk_h^2}{\omega^2}+\frac{1}{H_*}\right) \dP \; .
\end{eqnarray}

To see the magnitude of the spurious terms, we nondimensionalize,

\begin{eqnarray}
    \tilde{u} &=& - \frac{\tilde{\rho}_0^{-1}}{1-\eps^4+\eps^2\tilde{H}_*^{-1}}\left(\frac{d}{d\zeta} + \eps^3\right) \delta\tilde{P} \; ,
\\
    \tilde{w} &=& \frac{i\eps ~\tilde{\rho}_0^{-1}}{1-\eps^4+\eps^2\tilde{H}_*^{-1}}\left(1 + \eps \frac{d}{d\zeta} + \frac{\eps^2}{\tilde{H}_*}\right) \delta\tilde{P} \; ,
\end{eqnarray}

\noindent If one compares these expressions with Equations~\eqnref{eqn:nondim_u} and \eqnref{eqn:nondim_w}, it is clear that all spurious terms appear at second order in the dimensionless frequency. Since all of the other fluid variables (i.e., $\rho_1$, $P_1$, and $s_1$) are linear combinations of the two velocity components---see Equation~\eqnref{eqn:thermo_variables}, the wave functions for all of the fluid variables are correct to first order when the pseudo-incompressible approximation is utilized.

Both of the anelastic approximations falter. For the fiducial anelastic approximation the nondimensional forms for the two velocity components are

\begin{eqnarray}
    \tilde{u} &=& - \frac{\tilde{\rho}_0^{-1}}{1-\eps^4+\eps^2\tilde{H}^{-1}}\left(\frac{d}{d\zeta} - \eps\tilde{N}^2 + \eps^3\right) \delta\tilde{P} \; ,
\\
    \tilde{w} &=& \frac{i\eps ~\tilde{\rho}_0^{-1}}{1-\eps^4+\eps^2\tilde{H}^{-1}}\left(1 + \eps \frac{d}{d\zeta} + \frac{\eps^2}{\tilde{H}_*}\right) \delta\tilde{P} \; ,
\end{eqnarray}

\noindent and for the LBR anelastic formulation we obtain

\begin{eqnarray}
    \tilde{u} &=& - \tilde{\rho}_0^{-1} \tilde{\alpha}_{\rm LBR}^{-1} \left(\frac{d}{d\zeta} - \eps\tilde{N}^2 + \eps^3\right) \delta\tilde{P} \; ,
\\
    \tilde{w} &=& i\eps ~\tilde{\rho}_0^{-1} \tilde{\alpha}_{\rm LBR}^{-1} \left(1 + \eps \frac{d}{d\zeta} + \frac{\eps^2}{\tilde{H}}\right) \delta\tilde{P} \; ,
\end{eqnarray}

\noindent with $\tilde{\alpha}_{\rm LBR} \equiv 1-\eps^4 +\eps^2\tilde{N}^2+\eps^2\tilde{H}^{-1}$. Both have errors in the horizontal velocity that appear at first order (i.e., the term involving $\eps \tilde{N}^2$). The fiducial anelastic approximation has the added shame that the Lagrangian pressure fluctuation itself is only correct to zero order and hence all fluid variables suffer from the same deficiency. For the LBR approximation, the first order error in the horizontal velocity $u$ propagates to errors of similar size in the fluctuations of the Eulerian pressure $P_1$ and density $\rho_1$.


\section{Discussion}
\label{sec:discussion}

We have demonstrated that internal gravity waves within a fully-compressible fluid become pseudo-incompressible in the low-frequency limit. Discrepancies from the solutions for a fully compressible fluid appear at second order in the non-dimensional frequency, i.e., the relative errors are ${\cal O}(\omega^2/gk_h)$. Conversely, the two anelastic approximations that we consider are inconsistent in the terms they neglect or retain in the continuity equation and vertical momentum equation. This inconsistency leads to errors in the wave functions that appear at first order, ${\cal O}(\omega/\sqrt{g k_h})$.  A summary of the fractional errors in the vertical wavenumber, envelope scale length, and in the eigenfunctions appears in Table~\ref{tab:table2}.

\tabletwo

These errors in the eigenfunctions arise from errors in either the local vertical wavenumber (the short spatial scale) or in the amplitude envelope of the oscillations (the long spatial scale)---see Tables~\ref{tab:table1} and \ref{tab:table2}. Many of the errors in the local dispersion relation explicitly require vertical variation in the atmospheric profiles of the density scale height and buoyancy frequency. Both \citet{Brown:2012} and \citet{Vasil:2013} explicitly considered isothermal atmospheres for which the scale heights and the characteristic frequencies are constants. So many of the errors identified here failed to materialize in those previous studies. \citet{Brown:2012} examined the behavior of internal gravity waves under the influence of three distinct anelastic treatments (including the LBR and fiducial anelastic formulations), and found that the LBR formulation suffered from the least deviation from the fully compressible result. Here we have demonstrated that the apparent success of the LBR approximation is only in reproducing the local dispersion relation. If one considers the wave functions directly, the LBR anelastic approximation fails at first order, just like fiducial anelastic.

\subsection{Conservation of Energy}

We can explore conservation of energy under each approximation by computing the vertical energy flux $F(z)$. Using Abel's Identity, as we did for a fully-compressible fluid in section~\ref{subsec:first_derivative}, we find a general expression for the energy flux that is valid for all three sound-proofing treatments,

\begin{equation}
    F(z) = - \frac{i\omega^3}{4\rho_0\alpha} {\cal W}\left\{\dP, \dP^*\right\}(z) \; .
\end{equation}

\noindent Each approximation generates a distinct form for $\alpha$ and has a different Wronskian because the coefficients of the first-derivative term in the respective ODEs differ.  

For the pseudo-incompressible equations, using Equations~\eqnref{eqn:PI_alpha} and \eqnref{eqn:PI_theta}, we find that the vertical energy flux is a constant function of height,

\begin{eqnarray}
    \nonumber
    {\cal W}\left\{\dP,\dP^*\right\}(z) &=& C \exp\left\{\int\frac{1}{\alpha_{\rm PI}\rho_0}\frac{d\left(\alpha_{\rm PI}\rho_0\right)}{dz}dz\right\}
\\
    &=& C \alpha_{\rm PI}\rho_0 \; , 
\\
    F_{\rm PI}(z) &=& -\frac{i\omega^3\, C}{4} = {\rm constant}  \; .   
\end{eqnarray}

\noindent Hence, energy is conserved.  It is interesting to note that we have not utilized the small parameter in this derivation of the energy flux. So, energy is conserved even when the low-frequency expansions have questionable validity because the dimensionless frequency is not small. 

Performing the same calculations for the two anelastic treatments reveals that the LBR formulation conserves energy (for the same reasons that the pseudo-incompressible equations do) and the fiducial anelastic equations lack energy conservation,

\begin{eqnarray}
    F_{\rm FA}(z) &=& -\frac{i\omega^3\, C}{4} \frac{\rho_{*0}}{\rho_0} =  \frac{i\omega^3\, C}{4} e^{s_0(z)/c_p}  \; .
\\
    F_{\rm LBR}(z) &=& -\frac{i\omega^3\, C}{4} = {\rm constant} \; . 
\end{eqnarray}

\noindent The vertical energy flux $F_{\rm FA}$ derived from the fiducial anelastic equations depends on the atmosphere's specific entropy density and, thus, in an atmosphere without adiabatic stratification the wave will deposit or extract energy as it travels.

\subsection{Applicability in Numerical Simulations}

In numerical simulations, it is hard to overstate the utility in converting the continuity equation from a parabolic prognostic equation to an elliptic PDE constraint, as is accomplished by both the anelastic and pseudo-incompressible approximations,

\begin{equation*}  
    \frac{\partial \rho}{\partial t} = - \grad\cdot\left(\rho \bvec{u}\right) \longrightarrow \left\{
        \begin{array}{c}
             \textrm{anelasticity} \\
             \grad\cdot\left(\rho_0 \bvec{u}\right) = 0\; , \\
             \\
              \textrm{pseudo-incompressibility}  \\
             \grad\cdot\left(P_0^{1/\gamma} \bvec{u}\right) = 0\; .
        \end{array} \right.
\end{equation*}

\noindent  In addition to removing sound waves and hence unthrottling the simulation's timestep, the imposition of constraints with this form allow the fluid velocity to be expressed using stream functions.  Of course, this reduces the number of variables that must be evolved from one time step to the next. However, this is done at the expense of increasing the spatial order of the now reformulated momentum equations in stream function form that is now devoid of any elliptic constraints.  This may demand auxiliary  boundary conditions on the streamfunctions that are not readily available. Moreover, if linear coupling in the system is treated as explicit in numerical time-stepping algorithms, it is known, specifically for spectral schemes, that the numerical accuracy of the scheme can be degraded at high resolutions. Fortunately, recent advances have shown that this degradation is avoided if linear couplings remain implicit at the expense of using fully coupled implicit time-stepping schemes \citep{kJ09,marti2016,kB20,bM21}.

In the derivation of the pseudo-incompressible condition above, two related assumptions are made. First, the Mach number, Ma, of the flows is small such that the advection timescale is much longer than a sound-crossing time for a typical flow structure.  This ensures that fluid motions are in a constant state of pressure equilibration---i.e., the Eulerian pressure fluctuation is small. Second, we have assumed that fluctuations in the potential density are small compared to that of the background state. This later assumption is self-consistent with low-Mach number flows. Notably, unlike the anelastic approximation discussed below, it does not restrict density fluctuations to be small compared to that of the background state. Finally, since we have ignored diffusive effects in the derivation of the pseudo-incompressible constraint, i.e., we have ignored $Q$ in Equation~\eqnref{eqn:alt_continuity}, we have made the further assumption that the P\'eclet number is large, ${\rm Pe} \gg 1$, such that the thermal diffusion timescale is long compared to the advective time scale. To summarize, for the pseudo-incompressible constraint to be valid, we must have the following ordering of timescales,

\begin{equation}
    \tau_{\rm sound} \ll \tau_{\rm adv} \ll \tau_{\rm diff} \;,
\end{equation}

\noindent  or equivalently in terms of nondimensional numbers

\begin{eqnarray}
    {\rm Ma} &\equiv& \frac{\tau_{\rm sound}}{\tau_{\rm adv}} = \frac{U}{c} \ll 1 \;, 
\\
    {\rm Pe} &\equiv& \frac{\tau_{\rm diff}}{\tau_{\rm adv}} = \frac{UL}{\kappa} \gg 1 \;,
\end{eqnarray}

\noindent where $U$ is a typical flow speed, $L$ is a typical length scale, and $\kappa$ is the thermal diffusivity.

The validity of the anelastic constraint requires the same assumption of low Mach number, ${\rm Ma} \ll 1$, but makes a different stricture on the effectiveness of thermal diffusion. Since, we must ignore Eulerian fluctuations of the mass density in the continuity equation, the equation of state dictates that, in addition to small pressure fluctuations, we must have small entropy or temperature fluctuations. In the convection zone of a star or planet, where the stratification is essentially adiabatic, entropy fluctuations are naturally small; anelasticity holds; and the anelastic and pseudo-incompressible conditions are equivalent.  However, in a region of stable stratification, the only way that the entropy or temperature fluctuations can remain small is if temperature homogeneity is diffusively maintained across flow structures \citep[see][]{bannon1996anelastic}. This requires that the thermal diffusion time is short compared to the advective time scale. Summarizing, anelasticity requires

\begin{equation}
    \tau_{\rm sound}, \tau_{\rm diff}  \ll  \tau_{\rm adv} \;,
\end{equation}

\noindent  or equivalently

\begin{equation}
    {\rm Ma} \ll 1 \; , \qquad  {\rm Pe} \ll 1 \; .  
\end{equation}

The limitation of low Mach number is easily met in many astrophysical and geophysical applications. Convection is sedate in the Jovian planets, in the Earth's interior, and in the deep layers of low-mass stars. Wave motions and circulations in the stably stratified regions of stars and planets are similarly often low Mach number. The requirements on the P\'eclet number are usually the more restrictive of the two assumptions.  For example, the thermal diffusion time in the Sun is typically millions of years; using the solar radius as the length scale, $L = R_\odot \approx 700$ Mm, and a thermal diffusivity appropriate for photon diffusion, $\kappa \sim 10^7~{\rm cm}^2~{\rm s}^{-1}$, we obtain $\tau_{\rm diff} \sim$ 16 Myr. If we consider the meridional circulation at the base of the Sun's convection zone and adopt a typical flow speed of 1 m s$^{-1}$, we obtain an advective timescale of 20 years, leading to a P\'eclet number of ${\rm Pe} \sim 10^6$.  Clearly, these motions are not anelastic; thermal diffusion cannot act rapidly enough to eliminate the temperature fluctions generated by advection. However, the motions do satisfy both of the requirements for pseudo-incompressibility, ${\rm Ma}\ll 1$ and ${\rm Pe} \gg 1$.  Although large P\'eclet numbers are often true from an astrophysical perspective, numerical simulations are often performed in regimes where ${\rm Pe}\sim{\cal O}(1)$. The anelastic approximation offers no resolution to this problem, but the pseudo-incompressible equations do. The restriction on the P\'eclet number ${\rm Pe}$ can be relaxed if the irreversible thermodynamic terms are retained,

\begin{equation}
    \grad \cdot \left(\rho_{*0} \bvec{u}\right) = \frac{\rho_{*0}}{\rho_0}\frac{Q}{c_p T_0} \; ,
\end{equation}

\noindent where $\rho_{*0}$ is the potential density of the background state. Of course, the retention of $Q$ will usually render a stream function formalism without the requirement of an elliptic constraint impossible.

Finally, we wish to note a final advantage of the pseudo-incompressible approximation over anelasticity.  While both sound-proofing schemes are well justified in a convection zone where the stratification is nearly adiabatic, if one wishes to simulate both stable and unstable regions in the same computational domain, the pseudo-incompressible approximation allows one to do so smoothly with a uniform treatment. The anelastic approximation will result in flows that violate the underlying assumptions of the approximation.


    We thank Lydia Korre and Rafael Fuentes for enlightening conversations about the pseudo-incompressible approximation. This work was supported by NASA through grants 80NSSC18K1125, 80NSSC19K0267 and 80NSSC20K0193 (BWH) and by NSF by grant DMS 2308338 (KJ).


\bibliography{references}

\end{document}